\documentclass{iopconfser}
\usepackage{amsmath}
\usepackage{booktabs} 
\usepackage{graphicx}
\usepackage{subcaption}
\usepackage{cite}
\usepackage{xurl}

\begin{document}

\title{Zero-dimensional multi-physics-constrained parameter design and optimization for advanced quasi-isodynamic stellarators}

\author{Ziyuan SUN$^{1}$,  Zixuan GUO$^{1}$,  Xianglin HAO$^{1*}$, Longjun QIN$^{2}$, Qian LIU$^{1}$ and Xiang TENG$^{3}$}

\affil{$^1$Yan Fusion (Shanghai) Technology Co., Ltd, Shanghai 200120, People’s Republic of China}
\affil{$^2$PKU Shenzhen Graduate School – Stonehill Technology Joint Laboratory for Fusion and New Energy}
\affil{$^3$College of Physical Science and Technology, Bohai University, Jinzhou 121013, People’s Republic of China}
\affil{$^4$Department of Basic Education and Research, Liaoning Vocational University of Technology, Jinzhou 121007, People's Republic of China}
\email{haoxl@yanfusion.com}

\begin{abstract}
A zero-dimensional (0D) multi-physics-constrained framework for parameter design and optimization of Stable Quasi-Isodynamic Designs (SQuIDs) is presented. Single- and multi-objective optimizations for three staged devices are carried out using an in-house stellarator 0D systems code: YF-1 for discharge demonstration, YF-2 for scientific break even, and YF-3 for a commercial demonstration plant. Pareto searches map the main design trade-offs across the three generations. The equal weight optima for YF-2 and YF-3 both lie in the electron-root favorable regime of the adopted root proxy: YF-2 recovers $Q_{phys} \sim 1$, while YF-3 reaches an ignited point at reactor scale. Future work will couple engineering feasibility and economic assessment modules for integrated plant evaluation.
\end{abstract} 

\section{Introduction}
\label{sec:intro}

\par Yan Fusion (Shanghai) Technology Co., Ltd., established in March 2025, aims at commercial controlled fusion energy. The company is currently pursuing engineering verification of prototype superconducting coils, physics optimization of advanced stellarators, and early conceptual engineering studies of its first experimental device. Near term road map comprises three stages: YF-1 for discharge demonstration, YF-2 for scientific break even, and YF-3 for a commercial demonstration power plant.

\begin{figure}[htbp]
  \centering
  \includegraphics[width=0.85\linewidth]{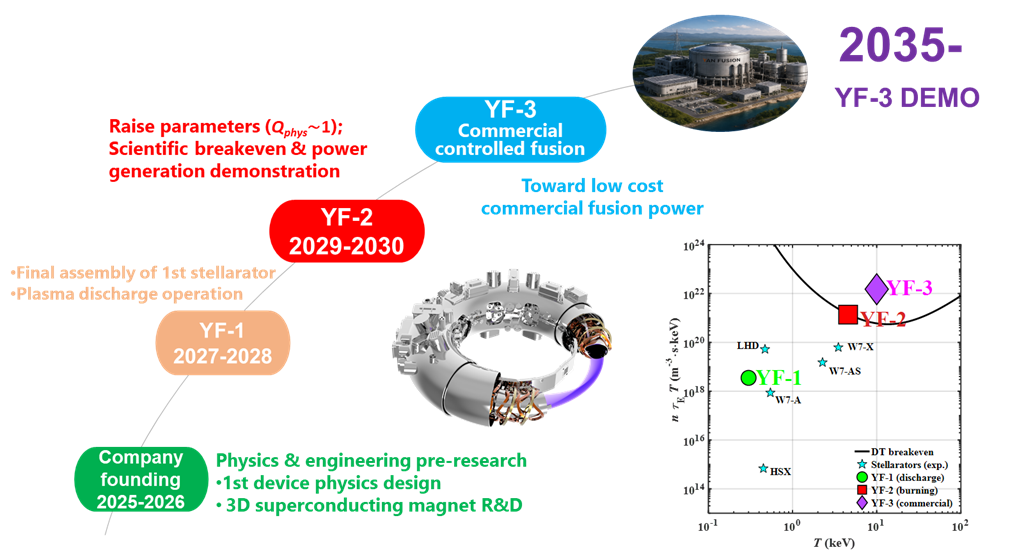}
  \caption{Yan Fusion development road map.}
  \label{fig:yf_roadmap}
\end{figure}

\par At the early design stage of such a staged program, the overall design point is still open. In particular, parameters such as the major radius $R_0$, the mean magnetic field $B_0$, the auxiliary heating power $P_{aux}$, and related quantities such as aspect ratio $A$, target fusion gain and operating density, among others, remain to be determined. High fidelity modeling, whether of essentially two dimensional tokamak equilibria or of fully three dimensional stellarator configurations, is indispensable for detailed design, but it is also computationally expensive: a single free boundary MHD equilibria reconstruction can require minutes to hours, and stellarator optimization campaigns commonly involve large ensembles of such evaluations~\cite{5merlo2021proof}. Data driven surrogate models have therefore been developed to accelerate three dimensional equilibrium calculations by orders of magnitude~\cite{5merlo2021proof,6merlo2023physics}. Their usefulness, however, remains bounded by the training domain, by the fidelity with which equilibrium proxies and global figures of merit are recovered, and by the need for substantial high fidelity data before the surrogate can be trusted outside its sampled subspace~\cite{5merlo2021proof,6merlo2023physics}. Consequently, surrogate acceleration alone does not replace the need for a compact, physics constrained scan that can explore device size, field strength and heating power before expensive three dimensional campaigns are launched.

\par Zero dimensional (0D) systems representations address precisely this gap. By compressing confinement, radiation, density limits, fusion gain and engineering envelopes into a small set of coupled constraints, they yield indicative ranges of plasma and machine parameters that are suitable for coarse sizing and for early discrimination among design options. This philosophy follows recent stellarator systems studies. System code PROCESS~\cite{9kovari2014process} has been developed with HELIAS specific modules for stellarator systems analysis~\cite{10warmer2015helias} and later extended to general stellarator power plants by ingesting reduced parameters extracted from three dimensional equilibria and coil sets~\cite{7lion2021general}, while high field stellarator studies have shown how 0D parameter scans can expose cost and performance trends before detailed engineering optimization~\cite{8prost2024economically}. Such analyses should not be read as quantitative predictions of a final machine. Absolute estimates remain indicative because profile physics, turbulent transport, coil geometry, divertor performance and cost assumptions are compressed into reduced models. Their value is comparative: they identify robust trends with magnetic field, aspect ratio, confinement improvement, density limit, auxiliary heating and fusion gain.

\par The present work is motivated by recent progress on Stable Quasi-Isodynamic Designs (SQuIDs), a class of advanced quasi-isodynamic (QI) stellarators proposed as reactor candidates~\cite{1goodman2024quasi,2goodman2025quasi}, together with related QI oriented optimization studies~\cite{3helander2024optimised,4neto2025electron}. We construct a 0D multiphysics constrained parameter design and optimization framework for an advanced QI stellarator under development at Yan Fusion. With the design point still open, the framework provides an overall estimate of plasma operating parameters, performs systems level optimization, and maps parameter sensitivities for coarse determination of machine size, heating requirements and accessible operating windows. The resulting operating maps are intended to guide the pre-selection of magnetic configurations and to furnish a transparent baseline for subsequent three dimensional equilibrium optimization, coil design and transport calculations. Future work will compare these 0D results with established systems codes and higher-fidelity analyses; the 0D layer reported here is therefore positioned as the first design step that underpins, rather than substitutes for, those three dimensional studies.

\par Remainder of this paper is organized as follows. Sec.~\ref{sec:2_0d_simu} presents the 0D physics model. Sec.~\ref{sec:3_simu_res} reports the 0D scan results. Sec.~\ref{sec:4_Opt_res} discusses optimization results and analysis for the three device generations. Conclusions are given in the final Sec.~\ref{sec:5_conclusion}, with supporting material collected in Appendix~A.

\section{Zero-Dimensional Simulation model}
\label{sec:2_0d_simu}
\par The 0D model closes a design point from a small set of global inputs. Given $R_0$, $B_0$, $A$, $ \left\langle \beta \right\rangle$, Sudo operating fraction $f_{sudo}$, the profile peaking exponents, and the confinement multiplier H, the model returns averaged density and temperature, loss power, fusion power, auxiliary power, fusion gain, etc. The approach follows the systems-analysis practice used in stellarator extensions of PROCESS~\cite{7lion2021general} and in recent high-field stellarator parameter studies~\cite{8prost2024economically}: global constraints are retained, and the radial profiles remain parametric.

\subsection{Plasma volume and surface area}
\label{sec:2.1_plasma_vol}

\par Two geometry options are available. The first, analytic option uses an elongated toroidal proxy: 
\begin{equation}
    \label{eq:01_tor_proxy}
    V_p = 2\pi^2 R_0 a^2 \kappa, 
\end{equation}
with elongation $\kappa$ held fixed. The same reduced description supplies a plasma surface estimate for wall loading calculations with a prescribed plasma–wall gap.

\par The second option reads volume and surface area from three-dimensional QI equilibria computed with DESC or VMEC. Reference values $\hat{V}$ and $\hat{S}$ at reference major and minor radii are scaled when $R_0$ and $A$ change, following the reduced parameter ingest used for general stellarator PROCESS studies~\cite{7lion2021general}. Aspect ratio remains an independent design variable. In this way the 0D map stays attached to a realistic QI shape, while free boundary MHD is evaluated only for the reference equilibria.

\subsection{Profiles, density limit and power balance}
\label{sec:2.2_profile_dens}

\par Electron density and temperature (with $T_i=T_e$) are prescribed as
\begin{equation}
n_e(x) = n_{e0}\left(1 - x^2\right)^{S_n}, \quad
T_e(x) = T_{e0}\left(1 - x^2\right)^{S_T}, \quad x = r/a,
\label{eq:02_e_densT}
\end{equation}
where the volume averages then satisfy $\langle n \rangle = n_{e0}/(1 + S_n)$ and $\langle T \rangle = T_{e0}/(1 + S_T)$. The stored thermal energy $W$ is computed from these shapes. $S_n$ and $S_T$ are peaking parameters within a fixed profile family.

\par The line averaged density follows the stellarator density scaling of Sudo et al.~\cite{11sudo1990scalings},
\begin{equation}
\bar{n}_{\text{sudo}}\left[10^{20}\,\text{m}^{-3}\right] = 0.25\, f_{\text{sudo}} P_{in}^{1/2} B_0^{1/2} / a R_0^{1/2},
\label{eq:03_sudo}
\end{equation}
and the temperature is obtained from fixed closure,
\begin{equation}
\langle \beta \rangle = \frac{4 \mu_0 \langle n \rangle e \langle T \rangle}{B_0^2},
\label{eq:04_beta_T}
\end{equation}
with an optional correction for the assumed peaking. The thermodynamic state is fixed by this design-point closure together with the confinement law below.

\par The energy confinement time is written as $\tau_E = H\tau_{scaling}$, with $\tau_{scaling}$ selectable among ISS04~\cite{12yamada2005characterization} and the Murari shear/shearless scalings~\cite{13murari2021scaling}:
\begin{equation}
\label{eq:05_tau_scaling}
\begin{aligned}
\tau_E &= H \times \tau_{scaling}\left(P_h, n_e, B_0, R_0, a, \iota_{2/3}\right) \\
&= H \times \begin{cases}
\tau_E^{ISS04} \\
\tau_E^{SR\,\text{Shear}} \\
\tau_E^{SR\,\text{Shearless}}
\end{cases} \\
&= H \times \begin{cases}
0.134 a^{2.28} R^{0.64} P_h^{-0.61} n_e^{0.54} B^{0.84} \iota_{2/3}^{0.41} \\[1.5ex]
\left(7.92_{7.73}^{8.11}\right) \times 10^{-2} \cdot a^{2.53_{2.48}^{2.58}} 2.48 R^{0.97_{0.94}^{1.01}} P_h^{-0.60_{-0.62}^{-0.58}} n_e^{0.45_{0.43}^{0.48}} B^{0.67_{0.63}^{0.70}} \iota_{2/3}^{0.50_{0.49}^{0.51}} \left( 1 + \mathrm{e}^{-\left(\frac{\left(\frac{a}{R}\right) - 0.19_{0.18}^{0.20}}{0.017_{0.015}^{0.022}}\right)^2} \right) \\[2.5ex]
\left(11.23_{10.69}^{11.78}\right) \times 10^{-2} a^{2.24_{2.23}^{2.25}} R^{0.64_{0.63}^{0.65}} P_h^{-0.66_{-0.67}^{-0.65}} n_e^{0.57_{0.57}^{0.58}} B^{1.03_{0.98}^{1.08}} \iota_{2/3}^{0.37_{0.35}^{0.38}} \frac{1}{1 + 3.78_{3.43}^{4.14} \mathrm{e}^{-\frac{2R}{R_{Av}}}}
\end{cases} .
\end{aligned}
\end{equation}
\par The multiplier $H$ measures confinement relative to the chosen base scaling. In Eq.~(\ref{eq:05_tau_scaling}) the heating power entering $\tau_{scaling}$ is denoted heating power $P_h$. Under steady state power balance, heating power equals the plasma loss power, $P_{h}=P_{loss}$, and after convergence of the outer iteration used for Eq.~(\ref{eq:03_sudo}), $P_{in}=P_{loss}$.
\par $P_{loss}$ is obtained from a self-consistent energy balance. In the radiation coupled mode used for the baseline maps,
\begin{equation}
\label{eq:06_Ploss}
P_{loss} = \frac{W}{\tau_E} + P_{brem},
\end{equation}
so that volumetric radiation enters the same equation that determines $\tau_E$. In the present work, line radiation $P_{line}$ is neglected and only bremsstrahlung is retained in $P_{loss}$.

\subsubsection{Deuterium--tritium operation}\label{sec:2.2.1_DD_op}\hfill\\
For a 50:50 D--T mixture, the fusion power is obtained by integrating the local thermal fusion reaction rate over the plasma,
\begin{equation}
\label{eq:07_Pfus}
P_{\mathrm{fus}} = E_{DT} \int n_D(x)n_T(x) \left\langle \sigma v \right\rangle_{DT}\!\left[T(x)\right] \,\mathrm{d}V,
\end{equation}
where $E_{DT}=17.59~\mathrm{MeV}$ is the energy released for each D--T reaction and $\left\langle \sigma v \right\rangle_{DT}$ is evaluated using the Bosch--Hale parametrization~\cite{14bosch1992improved}.
\par The fraction of the fusion power carried by alpha particles is $f_{\alpha}=3.5/17.59$, and the required auxiliary heating power and the fusion gain are
\begin{equation}
\label{eq:08_Paux}
P_{\mathrm{aux}} = P_{\mathrm{loss}}-f_{\alpha}P_{\mathrm{fus}},
\qquad
Q=\frac{P_{\mathrm{fus}}}{P_{\mathrm{aux}}}.
\end{equation}

On an ignited branch with $P_{\mathrm{aux}}\leq 0$, the gain is reported as formally unbounded in the usual systems-code sense. Quasi-neutrality includes deuterium, tritium, helium ash, and prescribed impurities. Helium ash is closed using the residence time $\tau_{\mathrm{He}}^{*} = \alpha_{\mathrm{He}} \tau_E$ and is iterated together with bremsstrahlung when radiative losses are retained in $P_{\mathrm{loss}}$.

\subsubsection{Deuterium–deuterium operation}\label{sec:2.2.2_DD_op}\hfill\\
A parallel D–D branch is retained for fuel-cycle comparisons and for early sizing of a pure-deuterium option. The primary reactions are the two D–D branches of approximately equal branching ratio,
\begin{equation}
\label{eq:09_DD_reation1}
\text{D} + \text{D} \rightarrow \text{T} + \text{p} \ (4.04\,\text{MeV}),
\end{equation}
\begin{equation}
\label{eq:10_DD_reation2}
\text{D} + \text{D} \rightarrow {}^{3}\text{He} + \text{n} \ (3.27\,\text{MeV}).
\end{equation}

Thermal reactivities for both branches are also taken from the Bosch–Hale parametrization~\cite{14bosch1992improved}. The total primary fusion power is obtained from a profile integral of the D–D rate, with a charged-particle heating fraction that averages the deposited energies of the two branches. When secondary burn is enabled, the primary tritium and ${}^{3}\text{He}$ ash drive the follow on channels:
\begin{equation}\label{eq11}
\text{D} + \text{T} \rightarrow {}^{4}\text{He} + \text{n} \ (17.59\,\text{MeV}),
\end{equation}
\begin{equation}\label{eq12}
\text{D} + {}^{3}\text{He} \rightarrow {}^{4}\text{He} + \text{p} \ (18.35\,\text{MeV}),
\end{equation}
The steady ash fractions of T, ${}^{3}\text{He}$ and ${}^{4}\text{He}$ are closed self-consistently: each species obeys a production–burn–exhaust balance with its own residence time $\tau^*_s = \alpha_s \tau_E$, and quasi-neutrality is enforced among $\text{D}$, the three ash species and the prescribed impurities. The total fusion power is then $P_{fus} = P_{\text{DD}} + P_{\text{DT}} + P_{\text{D}{}^3\text{He}}$, and plasma heating power uses the charged fractions of each channel in place of the single D–T factor $f_{\alpha}$. At the $\text{keV}$ temperatures of the present maps the absolute D–D power remains far below the D–T branch; the module is kept so that fuel choice, ash dilution and secondary burn stay explicit in the systems scan.
\subsection{Bremsstrahlung and the treatment of other radiation channels}
\label{sec:2.3_brems}
\par Bremsstrahlung is retained as the volumetric radiation loss in $P_{loss}$. The volume-integrated power uses a profile-aware thermal fit with relativistic corrections in the classical electron–ion and electron–electron lineage discussed by Svensson~\cite{15svensson1982electron}. Impurity charge enters through $Z_{eff}$ from the quasi-neutral mix; prescribed impurity fractions are already included in the 0D design and optimization.
\par Line radiation losses are neglected in $P_{loss}$ for the design points considered here, relative to bremsstrahlung and the conductive/convective loss $W/\tau_E$. A full line-radiation evaluation would require ionization balance and divertor conditions beyond the present core closure.
\par Synchrotron radiation losses are likewise assumed negligible compared with bremsstrahlung and the diffused power, following the treatment used in recent high field stellarator systems studies~\cite{8prost2024economically}. A Trubnikov type estimate of synchrotron loss~\cite{16trubnikov1979universal}, with later refinements for optical depth and wall reflection~\cite{17albajar2002electron,18albajar2009raytec}, can be used to check this assumption for selected design points; it is not included in the baseline $P_{loss}$ of the present maps.
\par With only bremsstrahlung retained in $P_{loss}$, a separate radiative density margin is monitored against the line-averaged radiation limit of W7-AS class stellarator operation, as recalibrated by Prost and Volpe~\cite{8prost2024economically}. The critical density is:

\begin{equation}
\label{eq:13_crit_dens}
n_{crit}\left[10^{20}\,\text{m}^{-3}\right] = C_c \left(\frac{P_h}{V_p}\right)^{0.48} B_0^{0.54}, \ C_c = 1.46,
\end{equation}
and a dimensionless operating factor is formed as $\lambda_{rad} = \langle n \rangle / n_{crit}$. The design scans and staged optimizations impose $\lambda_{rad} \le 1.5$. This margin flags density–power combinations outside an accepted radiative corridor. Divertor radiation distributions are left for later engineering assessment.
\subsection{Neutron wall loading}
\label{sec:2.4_n0_wall}

\par For D–T points the neutron wall load is estimated from the non-alpha fusion power divided by first wall area,
\begin{equation}
\label{eq:14_n0_wall_load}
N_{wall} = \frac{P_n}{A_{wall}} = \frac{(1 - f_\alpha) P_{fus}}{A_{wall}},
\end{equation}
when volume and surface are taken from a scaled QI equilibrium, the plasma surface is:
\begin{equation}
\label{eq:15_plasma_surf}
S = \hat{S} \left( \frac{R_0}{\hat{R}} \right) \left( \frac{a}{\hat{a}} \right), \quad a = \frac{R_0}{A},
\end{equation}
so that fixed $A$, $S \propto R_0^2$. The first wall area is obtained by expanding the plasma surface across a prescribed plasma–wall gap $g$,
\begin{equation}
\label{eq:16_wall_area_01}
A_{wall} = \frac{S(a + g)}{a}.
\end{equation}

\par In the analytic geometry option the same gap enters an elongated-torus wall proxy,
\begin{equation}
\label{eq:17_wall_area_02}
A_{wall} = \left(4\pi^2 A_d \kappa^{0.65} - 4\kappa\delta\right)(a + g)^2,
\end{equation}
with $A_d = R_0/(a+g)$. Divertor localization and peaking factors remain unresolved at this stage, so the reported wall load is a core averaged indicator for early sizing.

\subsection{Electron-Cyclotron and EBW Heating Windows, and Status of NBI}
\label{sec:2.5_heating_win}
\par Auxiliary heating in the present scans is assessed through electron-cyclotron (EC) accessibility. When the core density exceeds the O-mode or X-mode cutoff, coupling to electron Bernstein waves (EBW) through O–SX–B conversion is examined. Cutoff densities for O and X modes follow from B0 and from the central density implied by $S_n$. Overdense feasibility is decided by the cold-plasma O–SX–B window summarized in Appendix A. That appendix brackets accessible heating regions on the 0D map. Antenna design and full-wave absorption are outside its scope. 
\par Neutral-beam injection, including negative ion NNBI for high energy beams, is left outside the present 0D heating module. A systems level NBI model would need beam energy and species, shine through and charge exchange losses, fast ion orbit loss in three dimensional fields, and a port-and-duct geometry consistent with the QI coil set. The systems code PROCESS already handles reduced NBI bookkeeping (penetration, shine-through, and orbit-loss fractions) once these inputs are prescribed~\cite{7lion2021general,9kovari2014process}, while quantitative stellarator deposition and loss patterns require Monte-Carlo orbit tools such as BEAMS3D validated on W7-X~\cite{19lazerson2021modeling}. Those ingredients are not yet coupled to this core closure. For the staged YF programme the immediate question on the POPCON is whether an EC or EBW window exists at the target density and field; that question is answered by the EC/EBW module above. NBI power and current-drive estimates will be added once a narrower band of QI equilibria and port layouts is available.

\subsection{Electron root proxy}
\label{sec:2.6_ele_root_proxy}
\par In stellarators, the radial electric field $E_r$ is usually not prescribed externally, but is determined by the plasma through ambipolar neoclassical transport~\cite{20helander2014theory}:
\begin{equation}
\label{eq:18_E_rad}
\sum_{s} Z_s e \Gamma_s^{neo}(E_r) = 0,
\end{equation}
where $\Gamma_s^{neo}$ is the neoclassical radial particle flux of species $s$.

\par The electron root is the ambipolar solution branch for which $E_r > 0$ in the plasma core~\cite{3helander2024optimised,4neto2025electron}. It is attractive for reactor operation because it weakens neoclassical accumulation of high Z impurities and can assist their outward exhaust~\cite{4neto2025electron,23biglari1990diamond}. A transition from the core electron root to an outer ion root may also produce strong $\boldsymbol{E} \times \boldsymbol{B}$ shear, with a possible reduction of turbulent radial transport~\cite{1goodman2024quasi,3helander2024optimised}.

\par A quantitative prediction of $E_r$ requires a radially resolved ambipolar transport calculation. Such calculations evaluate the species particle fluxes and solve the ambipolarity condition self-consistently. Such calculations are beyond the intended fidelity of our model. The $C_{root}$ contour is only a design indicator and does not replace drift-kinetic neoclassical evaluations such as DKES, NEO-2, NTSS, or radial transport analyses.

\par For 0D design scans, an algebraic electron root proxy is introduced to capture the dominant trapped-electron collisionality dependence, based on the standard $1/\nu$ scaling,
\begin{equation}
\label{eq:19}
\nu_e^* \propto \frac{\nu_{ei}}{\omega_b} \propto \frac{Z_{eff} n_{e0} R_0}{T_{e0}^2},
\end{equation}
Accordingly, the electron root proxy $C_{\text{root}}$ is defined as
\begin{equation}
\label{eq:20_Croot}
C_{\text{root}} = \frac{Z_{eff} n_{e0}\left[10^{20}\right] R_0}{T_{e0}^2},
\end{equation}
where $Z_{eff}$ is obtained from the self consistent helium ash and the impurity model. Smaller $C_{\text{root}}$ indicates lower effective collisionality and therefore facilitate access to the electron-root branch.

\par In 0D parameter scans below, points with $C_{\text{root}} < C_{\text{crit}}$ are considered as electron root favorable, and the remaining points as ion root. We use $C_{\text{crit}} = 0.25$, as an order unity calibration to the electron root accessibility trends reported for SQuID-X reactor configurations~\cite{1goodman2024quasi,4neto2025electron}. The contour $C_{\text{root}} = C_{\text{crit}}$ should be interpreted only as a fast design indicator on POPCON, and doesn’t replace calculations using DKES or NTX code for drift-kinetic neoclassical transport, and NEOPAX for radial transport~\cite{24van1989variational,25beidler2011benchmarking,26uwplasma_neopax_transport_physics_flux_models}.

\subsection{Staged 0D optimization and later engineering coupling}
\label{sec:2.7_0D_opt}
\par In addition to fixed grid POPCON scans, the same 0D core is embedded in a staged optimization layer for the YF-1, YF-2 and YF-3 design generations. Scalar searches and multi-objective Pareto searches share one evaluation path: each trial calls the 0D core, applies EC/EBW, radiative and other constraints, and returns stage dependent objectives such as auxiliary power, fusion gain, Lawson product, wall load and alpha margin. 

\begin{figure}[htbp]
  \centering
  \includegraphics[width=0.85\linewidth]{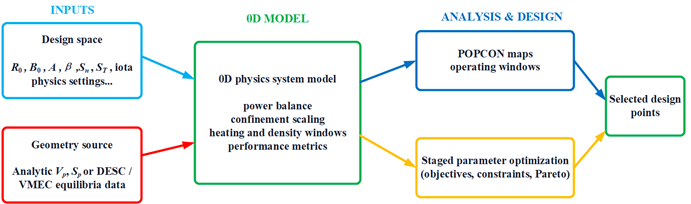}
  \caption{0D modeling and optimization framework.}
  \label{fig:0d_framework}
\end{figure}

\par The present paper is limited to physics-constrained sizing and staged optimization. Coupling to coil costing, plant power balance and economic modules, in the manner of systems studies~\cite{7lion2021general,8prost2024economically}, is planned once a narrower set of QI equilibria has been selected.

\section{0D Simulation results}
\label{sec:3_simu_res}
\par This section reports fixed-grid POPCON scans of the 0D core in Sec.~\ref{sec:2_0d_simu} on a reference four period QI equilibria. Volume and surface area are taken from the DESC reference and scaled with $R_0$ and $A$ as in Sec.~\ref{sec:2.1_plasma_vol}. Unless noted otherwise the maps use a 50:50 D–T mixture, $\langle \beta \rangle = 2\% $, $A = 4$, $f_{sudo} = 1.3$, confinement multiplier $H = 1.4$, temperature peaking $S_T = 0.5$, and helium ash residence $\tau_{He}^*=0.5\tau_E$. Dilute low Z impurities are prescribed at $f_C = 0.8\%$ and $f_O = 0.2\%$ relative to ion density $n_i$, which represent an optimistic wall and impurity mix for early sizing.

\subsection{D–T POPCON and EC/EBW domains}
\label{sec:3.1_DT_POPCON}

\par
\begin{figure}[htbp]
  \centering
  \begin{subfigure}[b]{0.48\textwidth}
    \centering
    \includegraphics[width=\textwidth]{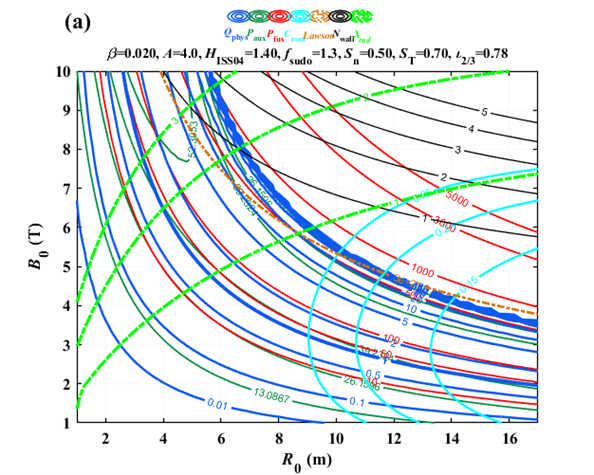}
    \caption{}
    \label{fig:popcon_a}
  \end{subfigure}
  \hfill
  \begin{subfigure}[b]{0.48\textwidth}
    \centering
    \includegraphics[width=\textwidth]{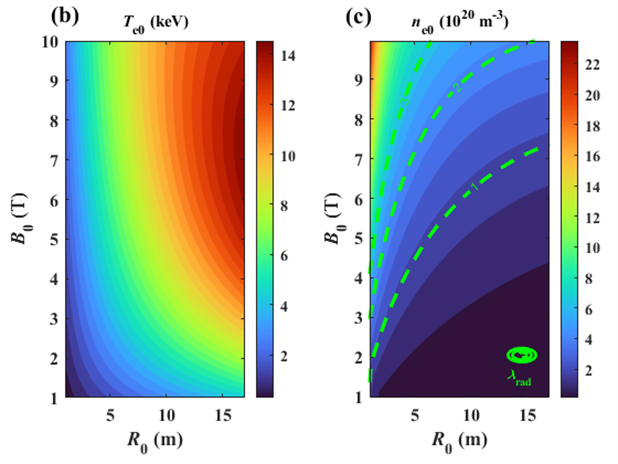}
    \caption*{(b, c)} 
    \label{fig:popcon_bc}
  \end{subfigure}
  \caption{POPCON results for $\langle\beta\rangle = 2\%$ and $A = 4$ with dilute C/O impurities ($f_{\text{C}} = 0.8\%$, $f_{\text{O}} = 0.2\%$) and $S_n = 0.50$: (a) performance and constraint contours in the ($R_0, B_0$) plane; (b) $T_{e0}$ and (c) $n_{e0}$.}
  \label{fig:popcon_results}
\end{figure}

 Fig.~\ref{fig:popcon_results} shows a representative D–T map at $S_n = 0.50$. Panel (a) collects the usual systems contours in $(R_0, B_0)$ plane: fusion gain $Q$, auxiliary power $P_{aux}$, fusion power$P_{fus}$, neutron wall load $N_{wall}$, radiative density factor $\lambda_{rad}$, and electron root proxy $C_{\text{root}}$. Panel (b) and (c) shows the corresponding central temperature $T_{e0}$ and density $n_{e0}$. Absolute levels remain indicative; the useful content is the relative placement of operating bands under the coupled density, beta, radiation and heating constraints.

\par
\begin{figure}[!ht]
  \centering
  \includegraphics[width=0.78\linewidth]{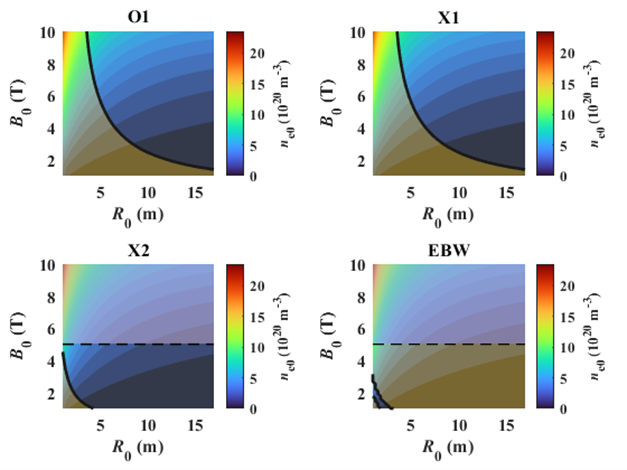}
  \caption{EC (O1/X1/X2) and EBW density feasibility domains in the
  $(R_0,B_0)$ plane at $S_n=0.50$ (same D--T closure as Fig.~\ref{fig:popcon_results}).}
  \label{fig:4}
\end{figure}

 Fig.~\ref{fig:4} overlays density feasibility for O1, X1, X2 and the O–SX–B/EBW channel at the same $S_n = 0.50$. Stellarator systems models like PROCESS enforce O-mode fundamental accessibility as a hard EC constraint~\cite{7lion2021general,10warmer2015helias}; X2 and EBW are shown here as comparative extensions on the same gyrotron frequency ceiling. An EBW feasible patch appears above the electromagnetic cutoffs: that region is the overdense extension discussed in Sec.~\ref{sec:3.3_dens_peaking}.

\subsection{D–D probe at fixed $(R_0, B_0)$} 
\label{sec:3.2_DD_probe}

\par The same QI geometry and closure switches admit a pure D–D fuel branch in Sec.~\ref{sec:2.2.2_DD_op}. To expose the fuel effect without remapping the whole plane, Tab.~\ref{tab:closure_comparison} compares D–T and D–D at one probe point, $R_0 = 2.80 \text{ m}$ and $B_0 = 2.20 \text{ T}$, with all other controls held fixed. At these temperatures the D–D fusion power is far below the D–T value, so Q and Lawson product drop sharply while $\tau_E$ changes only through the weaker $P_loss$ coupling. The probe is only a fuel cycle check.

\begin{table}[htbp]
\centering
\caption{Comparison of D--T and D--D closures at $R_0 = 2.80\text{ m}$, $B_0 = 2.20\text{ T}$.}
\label{tab:closure_comparison}
\begin{tabular}{ccccccc}
\toprule
Fuel & $\langle n \rangle (10^{20}\text{m}^{-3})$ & $\langle T \rangle (\text{keV})$ & $P_{loss} (\text{MW})$ & $P_{fus} (\text{MW})$ & $P_{aux} (\text{MW})$ & $Q_{phys}$ \\
\midrule
D--D & 0.95 & 1.16 & 4.75 & $\sim 0$ & 4.75 & $\sim 0$ \\
D--T & 0.95 & 1.16 & 4.75 & $1.68 \times 10^{-2}$ & 4.75 & $3.53 \times 10^{-3}$ \\
\bottomrule
\end{tabular}
\end{table}

\subsection{Density peaking $S_n$ and the EBW overdense extension} 
\label{sec:3.3_dens_peaking}

\par
\begin{figure}[!t]
  \centering
  \begin{subfigure}[b]{0.48\linewidth}
    \centering
    \includegraphics[width=\linewidth]{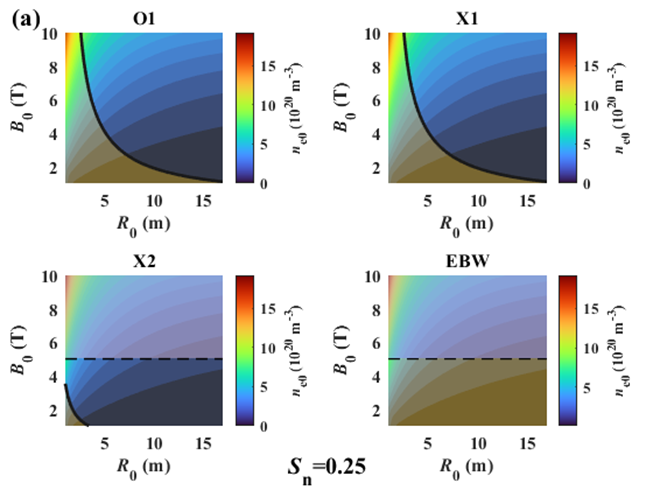}
    \caption{}
    \label{fig:ec_ebw_a}
  \end{subfigure}
  \hfill
  \begin{subfigure}[b]{0.48\linewidth}
    \centering
    \includegraphics[width=\linewidth]{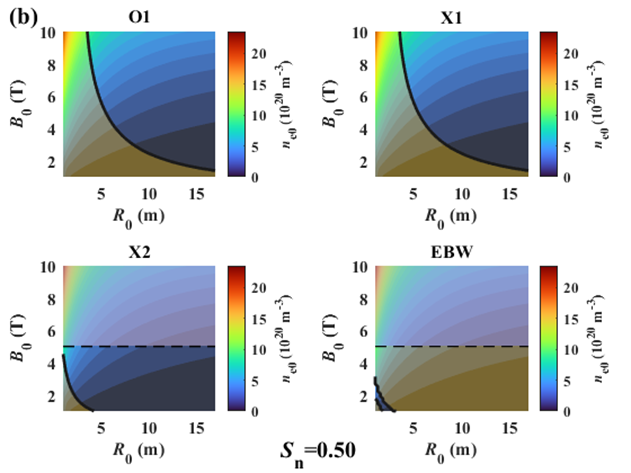}
    \caption{}
    \label{fig:ec_ebw_b}
  \end{subfigure}

  \vspace{0.8em} 

  \begin{subfigure}[b]{0.48\linewidth}
    \centering
    \includegraphics[width=\linewidth]{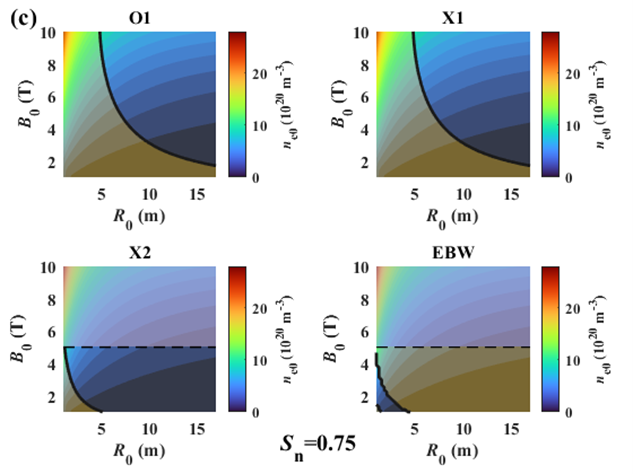}
    \caption{}
    \label{fig:ec_ebw_c}
  \end{subfigure}
  \hfill
  \begin{subfigure}[b]{0.48\linewidth}
    \centering
    \includegraphics[width=\linewidth]{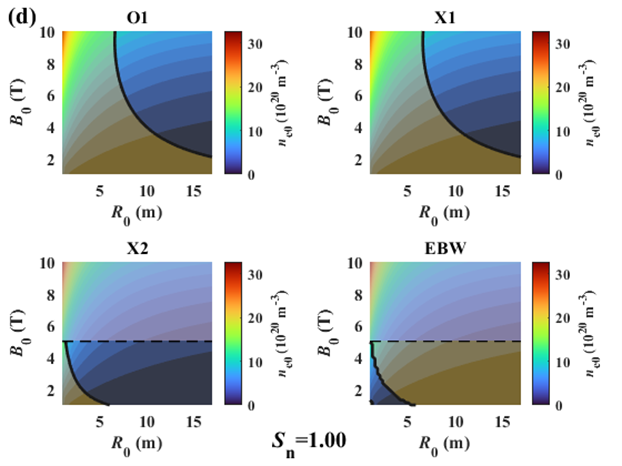}
    \caption{}
    \label{fig:ec_ebw_d}
  \end{subfigure}

  \caption{EC (O1/X1/X2) and EBW density feasibility in the ($R_0, B_0$) plane for (a) $S_n = 0.25$, (b) $0.50$, (c) $0.75$ and (d) $1.00$.}
  \label{fig:ec_ebw_feasibility}
\end{figure}

 Fig.~\ref{fig:ec_ebw_feasibility} repeats the EC/EBW density maps for $S_n = 0.25, 0.50, 0.75$ and $1.00$ at fixed $\langle \beta \rangle$, $A$ and $f_{sudo}$. Increasing $S_n$ raises the central density at fixed volume average, and steepens the profile used in the O–SX–B window. Higher $n_{e0}$ pushes more $(R_0, B_0)$ cells above the O1/X1 electromagnetic cutoffs, while the cutoff radius $r_{cut}$ and the density scale length $L_n$ at that layer shift with $S_n$. Small $S_n$ gives a flatter core and can be read loosely as H-mode-like flatness in systems senser.

\par On the O1 and X1 panels the infeasible area grows with $S_n$, as expected when $n_{e0}$ increases against a $B_0$ dependent cutoff. The EBW panel behaves differently because that channel is built to cover the overdense window $n_{O,cut} < n_{e0} < n_{hi}$ together with the conversion efficiency and radial constraints. Cells that have just crossed the O1/X1 cutoff are precisely the cells the EBW branch is meant to test. Hence the EBW feasible overdense footprint becomes more visible as $S_n$ rises. 

\par A second, weaker coupling comes from profile shape. At fixed $n_{e0}/n_{cut}$, larger $S_n$ moves $r_{cut}$ inward and shortens $L_n \propto 1/(S_nr)$  near the conversion layer. In Appendix~\ref{sec:appendix_a}, a shorter $L_n$ tends to help O–X transmission. Extreme peaking can also drive $r_{cut}$ toward the axis and distort $L_n$ enough to open a little infeasible pockets (visible near some corners of Figs.~\ref{fig:ec_ebw_c} and \ref{fig:ec_ebw_d}, so the EBW area does not grow monotonically with $S_n$.

\par The maps that mark an “extension relative to X2” need a careful reading. The EBW branch shares the same gyrotron frequency ceiling as the X2 electromagnetic branch, $\omega_{gyro}=\min (2\omega_{ce}, \omega_{\max})$, and the same $B \le B_{\max}$ resonance bound. Dashed guides on the EBW panel are aligned with the X2 panel for that reason: the scan is still keyed to the second-harmonic frequency capability. When the EBW branch is keyed to the X2 frequency ceiling, $n_{O,cut}$ denotes the O-cutoff density evaluated at that same $\omega = min(2\omega_{ce},\omega_{\text{max})}$, instead of an X-mode cutoff substitution. Below this floor the profile has no O-cutoff layer at the chosen $\omega$, so O→SX→B cannot start; above it the overdense EBW window opens. Power that reaches the plasma in that window is absorbed as Bernstein waves after O–SX–B conversion, not as an X2 electromagnetic branch pushed to higher density. In short, Fig.~\ref{fig:ec_ebw_feasibility} shows where X2 electromagnetic heating is cut off while O-SX–B/EBW may still couple under the same frequency budget: an EBW supplement above the X2 cutoff, in the spirit of EC bookkeeping in other stellarator models~\cite{7lion2021general} and of the cold-plasma O–SX–B window used here~\cite{27guo2017one}. Repeating the scan with cutoff mode O1 would draw the same kind of overdense extension against the fundamental O1 cutoff instead of X2.

\par Taken together, $S_n$ scan shows that peaking simultaneously lifts $n_{e0}$ and reshapes the conversion layer. O1/X1 electromagnetic access shrinks, while the EBW overdense channel occupies a larger share of the map. That trend is useful for early heating-port discussions on the YF POPCON, but it remains a 0D accessibility bracket.

\section{Optimization results}
\label{sec:4_Opt_res}
\par The 0D core of Sec.~\ref{sec:2_0d_simu} is embedded in a staged optimizer for the three YF device generations. Each trial evaluates the same closure and applies EC/EBW, radiative density and other stage dependent constraints. This section reports the optimized design points and parameter sensitivity scans.

\subsection{YF-1 optimization}
\label{sec:4.1_yf1}
\par YF-1 is the discharge demonstration stage. Fusion gain is not the design driver. The sole objective used here is to minimize the auxiliary heating power $P_{aux}$ that sustains the 0D point. A lower bound $Q_{phys,min}$ is not applied at this stage.

\par The free variables and box bounds follow the full YF-1 stage: $R_0 \in [2.2, 2.45] \text{ m}$, $B_0 \in [1.5, 3.0] \text{ T}$, $A \in [5.0, 11.0]$, $\langle \beta \rangle \in [0.01, 0.05]$, $S_n \in [0.2, 1.2]$, $S_T \in [0.2, 1.2]$, and $\iota_{2/3} \in [0.81, 0.90]$. The confinement multiplier and Sudo fraction are fixed at $H = 1.4$ and $f_{sudo} = 1.3$. Volume and wall geometry use the JSON file scaling of the 0D core.

\par A trial is feasible only when all of the following hold. (i) The 0D closure converges. (ii) ECRH access is satisfied in the X2 or EBW sense: the point lies in the X2 window or in the EBW O-SX-B overdense window (gyrotron frequency ceiling 280 GHz). (iii) The Prost radiative density constraint is met, $\lambda_{rad} \le 1.5$. (iv) $P_{aux} > 0$. These four filters are the nonlinear constraints passed to the optimizer. Density itself remains set by Sudo/ISS04 closure inside the 0D solve.

\par Tab.~\ref{tab:optimization_results_yf1} lists the YF-1 point returned by this $P_{aux}$ minimization. The coordinates are $R_0 = 2.26\ \text{m}$, $B_0 = 1.64\ \text{T}$ and $A = 11.00$, with $B_0R_0 =3.71\ \text{T}\cdot \text{m}$. The volume density average $\langle n\rangle = 1.43\times 10^{20}\ m^{-3}$ and $\langle T\rangle = 0.24\ \text{keV}$. The heating cost is $P_{aux} = 0.32\ \text{MW}$. Fusion power and $Q_{phys}$ are negligible at this temperature. The heating path is EBW (EC X2/EBW = 0/1). This point is the heating cost minimum inside the allowed window. Engineering layout is fixed later by a separate design loop.

\begin{table}[htbp]
\centering
\caption{Optimization results for YF-1: design variables and derived 0D outputs.}
\label{tab:optimization_results_yf1}
\begin{tabular}{cccccc}
\toprule
$R_0 (\text{m})$ & $B_0 (\text{T})$ & $B_0 R_0 (\text{T}\cdot\text{m})$ & $A$ & $\langle\beta\rangle$ & $S_n$ \\
\midrule
2.26 & 1.64 & 3.71 & 11.00 & 0.01 & 0.30 \\
\midrule
$S_T$ & $\iota_{2/3}$ & $P_{aux} (\text{MW})$ & $P_{fus} (\text{MW})$ & $Q_{phys}$ & $n_{e0} (10^{20}\,\text{m}^{-3})$ \\
\midrule
0.32 & 0.90 & 0.32 & $2.5\times 10^{-8}$ & $7.7\times 10^{-8}$ & 1.85 \\
\midrule
$\langle n\rangle (10^{20}\,\text{m}^{-3})$ & $T_{e0} (\text{keV})$ & $\langle T\rangle (\text{keV})$ & EC X2 / EBW & $\tau_E (\text{s})$ & $\frac{\langle n\rangle \tau_E \langle T\rangle}{(\text{m}^{-3}\cdot\text{s}\cdot\text{keV})}$ \\
\midrule
1.43 & 0.31 & 0.24 & 0/1 & 0.10 & $3.51\times 10^{18}$ \\
\bottomrule
\end{tabular}
\end{table}

\par
\begin{figure}[!t]
  \centering
  \begin{subfigure}[b]{0.48\linewidth}
    \centering
    \includegraphics[width=\linewidth]{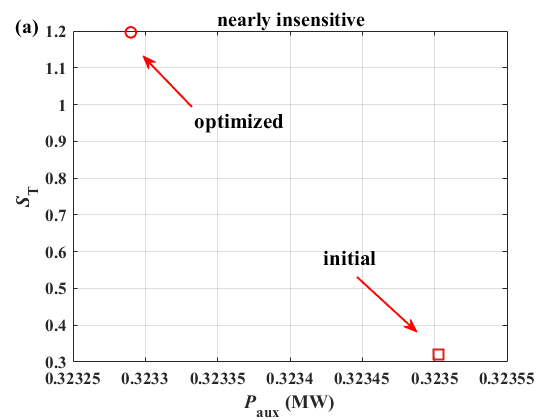}
    \caption{}
    \label{fig:one_para_scan_yf1_a}
  \end{subfigure}
  \hfill
  \begin{subfigure}[b]{0.48\linewidth}
    \centering
    \includegraphics[width=\linewidth]{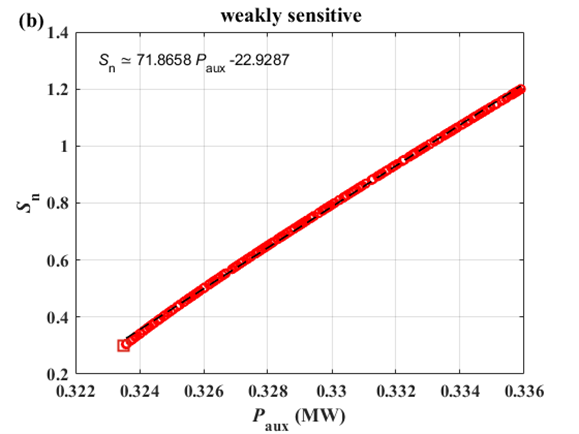}
    \caption{}
    \label{fig:one_para_scan_yf1_b}
  \end{subfigure}

  \vspace{0.8em} 

  \begin{subfigure}[b]{0.48\linewidth}
    \centering
    \includegraphics[width=\linewidth]{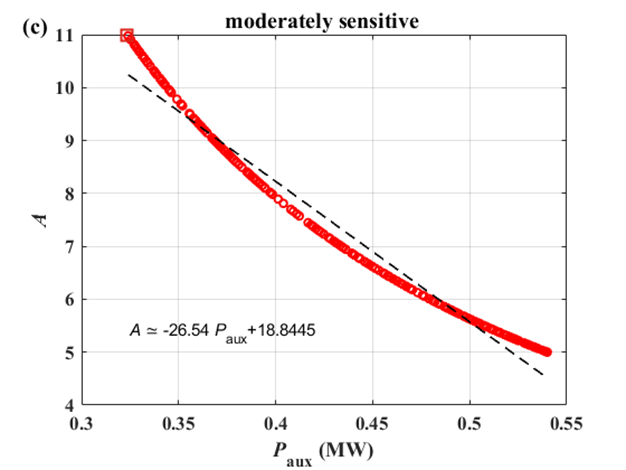}
    \caption{}
    \label{fig:one_para_scan_yf1_c}
  \end{subfigure}
  \hfill
  \begin{subfigure}[b]{0.48\linewidth}
    \centering
    \includegraphics[width=\linewidth]{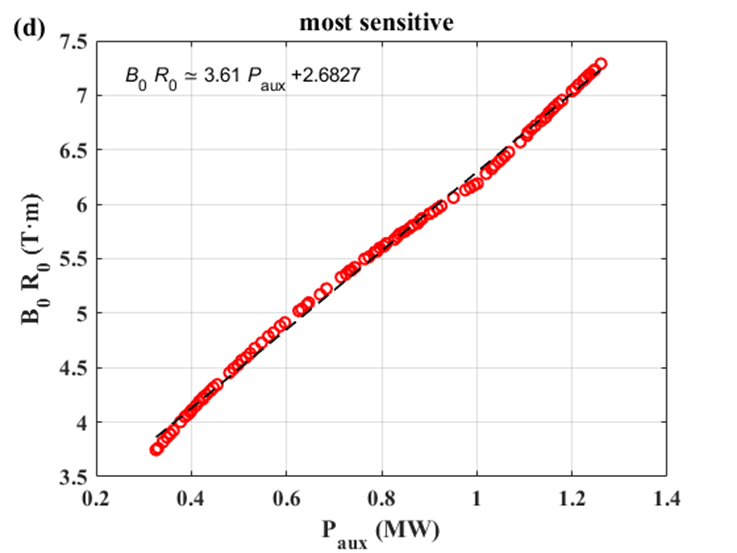}
    \caption{}
    \label{fig:one_para_scan_yf1_d}
  \end{subfigure}

  \caption{One parameter scans of $P_{aux}$ for YF-1 about the optimized seed: (a) $S_T$; (b) $S_n$; (c) $A$; (d)$ B_0R_0$. Other design variables are held fixed.}
  \label{fig:one_para_scan_yf1}
\end{figure}

 To rank the levers that set $P_{aux}$, we restart from this seed and vary one free variable at a time while the others are frozen. Fig.~\ref{fig:one_para_scan_yf1} shows the traces for $S_T$, $S_n$, $A$ and the size field product $B_0R_0$. $P_{aux}$ is nearly flat in $S_T$, weakly rising in $S_n$, moderately falling in $A$, and steeply falling in $B_0R_0$. Among these four scans, $B_0R_0$ dominates the heating budget.

\par The weak $S_T$ dependence follows from the fixed beta closure in Sec.~\ref{sec:2_0d_simu}. Once $\langle n\rangle$ and $B_0$ are set, the volume averaged temperature is largely fixed by $\langle \beta \rangle \propto \langle n \rangle T / B_0^2$.

\par Changing $S_T$ mainly reshapes $T_{e0}=\langle T \rangle (1+S_T)$. The stored energy $W$ and the ISS04 confinement time depend on volume averages. At the Tab.~\ref{tab:closure_comparison} point $P_{fus}$ is negligible beside $P_{aux}$, so profile driven changes in $P_{fus}$ feed back only weakly into the power balance. Hence $P_{aux}$ barely moves when $S_T$ is scanned about the seed.

\par $S_n$ enters more channels. The central density scales as $n_{e0}=\langle n\rangle(1+S_n)$, so a larger $S_n$ raises $n_{e0}$ at fixed volume average and shifts the ECRH/EBW feasibility test. The same exponent appears in the profile factor of $W$,
\begin{equation}
\label{eq:21_W_profile_factor}
W \propto \langle n \rangle \langle T \rangle (1 + S_n)(1 + S_T) / (1 + S_n + S_T),
\end{equation}
and in the O-SX-B layer used for overdense heating. Those links make $S_n$ a weak but visible driver of $P_{aux}$ on Fig.~\ref{fig:one_para_scan_yf1_b}.

\par Aspect ratio $A$ changes the minor radius $a$ at fixed $R_0$. Volume, surface area, and the Sudo density scaling and the ISS04 exponents in a all move together. A larger $A$ shrinks the plasma cross section and lowers the heating needed to hold the same beta class point. That geometric trend matches the moderate downward slope in Fig.~\ref{fig:one_para_scan_yf1_c}.

\par The product $B_0R_0$ sets the overall device scale and field. ISS04 confinement rises strongly with $B_0$ and $R_0$ (and with a at fixed $A$). At fixed density the beta closure raises $\langle T\rangle$ roughly as $B_0^2$. Both effects cut the required $P_{aux}$ once size and field increase. Fig.~\ref{fig:one_para_scan_yf1_d} therefore shows the steepest slope. On the present map, $B_0R_0$ is the primary heating cost lever for YF-1.

\par In summary, the YF-1 search minimizes $P_{aux}$ alone under EC and radiative filters. The returned point in Tab.~\ref{tab:optimization_results_yf1} is compact in heating cost, with negligible fusion gain. One parameter scans about that seed show $P_{aux}$ nearly insensitive to $S_T$, weakly sensitive to $S_n$, moderately sensitive to $A$, and most sensitive to $B_0R_0$. The ordering matches the fixed beta, Sudo and ISS04 structure of the 0D model.

\subsection{YF-2 optimization}
\label{sec:4.2_yf2}

\par The 0D core of Sec.~\ref{sec:2_0d_simu} is embedded in a staged optimizer for the three YF device generations. Each trial evaluates the same closure and applies ECRH/EBW, radiative density and other stage dependent constraints. We then report the YF-2 and YF-3 optimized points and the multi-objective Pareto scans.

\par YF-2 is the burning stage on a five period QI geometry scaled from the reference equilibrium. The design aim is a physical gain near unity, $Q_{phys} \sim 1$. The free variables are again $R_0$, $B_0$, $A$, $\langle \beta \rangle$, $S_n$, $S_T$ and $\iota_{2/3}$. The box bounds are $R_0 \in [6.0, 12.0] \text{ m}$, $B_0 \in [3.0, 8.0] \text{ T}$, $A \in [3.0, 11.0]$, $\langle \beta \rangle \in [0.01, 0.05]$, $S_n \in [0.2, 1.2]$, $S_T \in [0.2, 1.2]$ and $\iota_{2/3} \in [0.81, 0.90]$. $H = 1.4$ and $f_{sudo} = 1.3$ are held fixed.

\par The scalar cost used for a single SQP solve with equal weights is,
\begin{equation}
\label{eq:22_weights_yf2}
\text{F}_{YF-2} = w_{root}\hat{C}_{root} + w_{aux}\hat{P}_{aux} - w_{n\tau T}\left(n\tau_E T / L_{crit}\right),
\end{equation}
with $w_{root} = w_{aux} = w_{n\tau T} = 1$ and $L_{crit} = 3.0\times 10^{21}\ \text{m}^{-3}\cdot \text{s} \cdot \text{keV}$. Hats in the implementation are only scale factors so that the three terms sit at similar magnitude. The search therefore lowers $C_{\text{root}}$ and $P_{aux}$ while raising the volume Lawson product. A trial is considered feasible when the 0D closure converges, $0.95 \leq Q_{\mathrm{phys}} \leq 1.05$, and EC access is available through at least one permitted heating channel. YF-1 permits X2 or EBW access, whereas YF-2 and YF-3 also admit O1 access.

\par Tab.~\ref{tab:yf2_equal_weights} lists the equal weight SQP point. We find $R_0 = 6.00\ \text{m}$ and $B_0 = 4.70\ \text{T}$, with $A = 3.00$ at the lower edge of the box. The major radius is of the same order as present mid size devices such as LHD and W7-X. The aspect ratio is smaller than those machines. At this point $Q_{phys} = 1.05$, $P_{aux} = 13.6\ \text{MW}$ and $C_{\text{root}} = 0.20$. Following Sec.~\ref{sec:2.6_ele_root_proxy}, $C_{\text{root}} < 0.25$ is read as electron root favorable. The Lawson product is $1.37\times 10^{21}\ \text{m}^{-3}\cdot \text{s} \cdot \text{keV}$, still below $L_{crit}$. Heating is available on both O1 and X2 (EC O1/X2/EBW = 1/1/0).

\begin{table}[htbp]
\centering
\caption{YF-2 optimization results with equal objective weights: design variables and derived 0D outputs.}
\label{tab:yf2_equal_weights}
\begin{tabular}{cccccc}
\toprule
$R_0 (\text{m})$ & $B_0 (\text{T})$ & $A$ & $\langle\beta\rangle$ & $S_n$ & $S_T$ \\
\midrule
6.00 & 4.70 & 3.00 & 0.01 & 0.20 & 0.20 \\
\midrule
$C_{\text{root}}$ & $\iota_{2/3}$ & $P_{aux} (\text{MW})$ & $P_{fus} (\text{MW})$ & $Q_{phys}$ & $n_{e0} (10^{20}\,\text{m}^{-3})$ \\
\midrule
0.20 & 0.90 & 13.60 & 14.28 & 1.05 & 0.74 \\
\midrule
$\langle n\rangle (10^{20}\,\text{m}^{-3})$ & $T_{e0} (\text{keV})$ & $\langle T\rangle (\text{keV})$ & EC O1/X2/EBW & $\tau_E (\text{s})$ & $\frac{\langle n\rangle \tau_E \langle T\rangle}{(\text{m}^{-3}\cdot\text{s}\cdot\text{keV})}$ \\
\midrule
0.61 & 5.48 & 4.57 & 1/1/0 & 4.87 & $1.37\times 10^{21}$ \\
\bottomrule
\end{tabular}
\end{table}

\par
\begin{figure}[!t]
  \centering
  \begin{subfigure}[b]{0.48\textwidth}
    \centering
    \includegraphics[width=\textwidth]{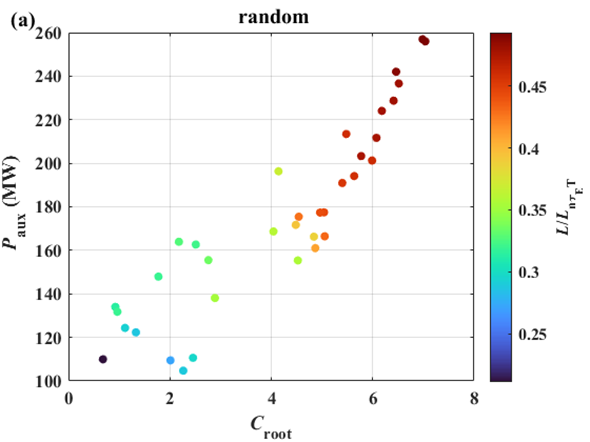}
    \caption{}
    \label{fig:Pareto_yf2_a}
  \end{subfigure}
  \hfill
  \begin{subfigure}[b]{0.48\textwidth}
    \centering
    \includegraphics[width=\textwidth]{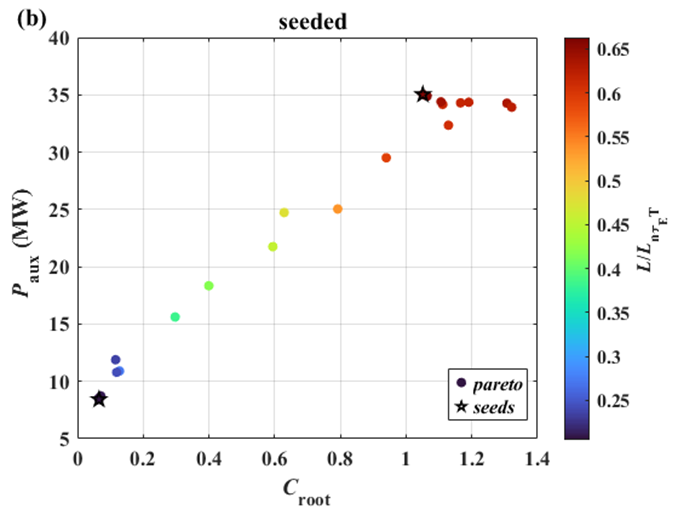}
    \caption{}
    \label{fig:Pareto_yf2_b}
  \end{subfigure}
  \caption{Multi-objective Pareto results for YF-2 in the $(C_{\text{root}},P_{aux})$ plane: (a) random initialization; (b) seeded initialization. Black markers denote the Pareto set; stars mark the seeds.}
  \label{fig:Pareto_yf2}
\end{figure}

 A three objective Pareto search is run on the same filters. The vector to be minimized is $\left(\hat{C}_{root}, \hat{P}_{aux}, -(n\tau_E T)/L_{crit}\right)$. Fig.~\ref{fig:Pareto_yf2} compares a random start with a seeded start that injects single objective anchors. Under random starts the cloud sits at higher $\hat{P}_{aux}$ and larger $\hat{C}_{root}$. With seeds the front reaches $C_{\text{root}}$ near 0.20 and $P_{aux}$ near $13.6\ \text{MW}$, consistent with the equal weight point in Tab.~\ref{tab:yf2_equal_weights}. The seeded front therefore recovers lower heating cost and stronger electron root tendency inside the $Q_{phys} \sim 1$ band.

\subsection{YF-3 optimization}
\label{sec:4.3_yf3}
\par YF-3 is the demo or reactor stage on the same five period QI scaling family. The free variables and fixed $H$, $f_{sudo}$ are as in Sec.~\ref{sec:4.2_yf2}. The box bounds widen to $R_0 \in [8.0, 17.0]\ \text{m}$, $B_0 \in [5.0, 10.0]\ \text{T}$ and $A \in [5.0, 11.0]$, with the same ranges for $\langle \beta \rangle$, $S_n$, $S_T$ and $\iota_{2/3}$. Ignited branches with $P_{aux} \le 0$ are admitted.

\par The scalar cost for an equal weight SQP solve is 
\begin{equation}
\label{eq:23_weights_yf3}
\text{F}_{YF-3} = w_{aux}\hat{P}_{aux} + w_{root}\hat{C}_{root} + w_N\hat{N}_{wall} - w_\alpha\hat{M}_\alpha,
\end{equation}
with all four weights also equal to 1. Here $N_{wall}$ is the neutron wall load and $M_\alpha = P_\alpha / P_{loss}$ is the alpha heating multiplier. On an ignited branch $P_{aux}$ is already near zero, so the useful trade is among $C_{\text{root}}$, $N_{wall}$ and $M_\alpha$. Feasibility requires 0D closure plus $Q_{phys} \ge 10$, $n\tau_E T \ge 3.0 \times 10^{21} \text{ m}^{-3}\cdot\text{s}\cdot\text{keV}$, $N_{wall} \le 5.0 \text{ MW/m}^{-2}$, EC access under O1 or X2 or EBW. $M_\alpha$ is maximized in the cost; a hard floor $M_\alpha \ge 1$ is optional and is off in the default run. Points with $C_{\text{root}} < 0.25$ remain electron root favorable by the Sec.~\ref{sec:2.6_ele_root_proxy} proxy.

\par Tab.~\ref{tab:yf3_equal_weights} lists the equal weight point. The machine is reactor scale: $R_0 = 17.00 \text{ m}$, $B_0 = 9.30 \text{ T}$ and $A = 5.00$. The solution is ignited, with $P_{aux} = 0$, $Q_{phys} \to \infty$ and $P_{fus} = 5.16 \text{ GW}$. We obtain $M_\alpha = 3.3$ and $C_{\text{root}} = 0.13$, well inside the electron root favorable band. The Lawson product is $1.50 \times 10^{22} \text{ m}^{-3}\cdot\text{s}\cdot\text{keV}$. Heating access is on O1 (EC O1/X2/EBW = 1/0/0).

\begin{table}[htbp]
\centering
\caption{YF-3 optimization results with equal objective weights: design variables and derived 0D outputs.}
\label{tab:yf3_equal_weights}
\begin{tabular}{cccccc}
\toprule
$R_0(\text{m})$ & $B_0(\text{T})$ & $A$ & $\langle\beta\rangle$ & $S_n$ & $S_T$ \\
\midrule
17.00 & 9.30 & 5.00 & 0.01 & 0.20 & 1.20 \\
\midrule
$C_{\text{root}}$ & $\iota_{2/3}$ & $M_\alpha$ & $P_{aux} (\text{MW})$ & $P_{fus} (\text{GW})$ & $Q_{phys}$ \\
\midrule
0.13 & 0.90 & 3.3 & 0(ignited) & 5.16 & $\infty$ \\
\midrule
$n_{e0} (10^{20}\,\text{m}^{-3})$ & $\langle n\rangle (10^{20}\,\text{m}^{-3})$ & $T_{e0} (\text{keV})$ & $\langle T\rangle (\text{keV})$ & EC O1/X2/EBW & $\frac{\langle n\rangle \tau_E \langle T\rangle}{(\text{m}^{-3}\cdot\text{s}\cdot\text{keV})}$ \\
\midrule
1.61 & 1.34 & 17.23 & 7.83 & 1/0/0 & $1.50\times 10^{22}$ \\
\bottomrule
\end{tabular}
\end{table}

\par
\begin{figure}[!t]
  \centering
  \begin{subfigure}[b]{0.48\textwidth}
    \centering
    \includegraphics[width=\textwidth]{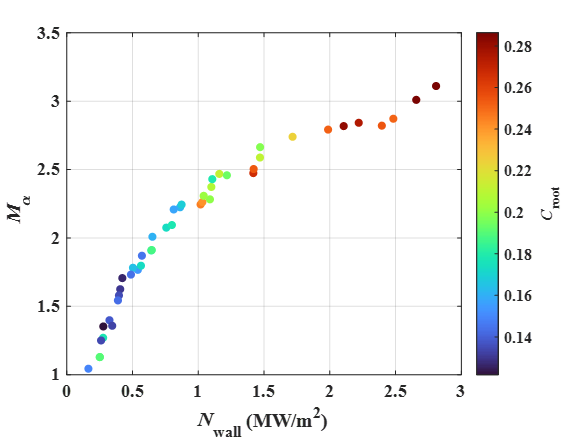}
    \caption{}
    \label{fig:Pareto_yf3_a}
  \end{subfigure}
  \hfill
  \begin{subfigure}[b]{0.48\textwidth}
    \centering
    \includegraphics[width=\textwidth]{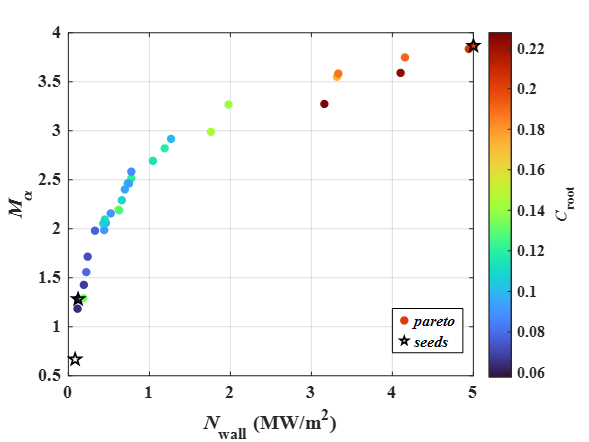}
    \caption{}
    \label{fig:Pareto_yf3_b}
  \end{subfigure}
  \caption{Multi objective Pareto results for YF-3 in the $(N_{wall}, M_{\alpha})$ plane: (a) random initialization; (b) seeded initialization. Color indicates $C_{\text{root}}$. Markers denote the Pareto set; stars mark the seeds.}
  \label{fig:Pareto_yf3}
\end{figure}

 The corresponding Pareto search minimizes $\left(\hat{C}_{root}, \hat{N}_{wall}, -\hat{M}_\alpha\right)$. $P_{aux}$ is dropped from the vector because it vanishes on the ignited branch. Fig.~\ref{fig:Pareto_yf3} again contrasts random and seeded starts. Random starts already trace a clear front in the $(N_{wall}, M_\alpha)$ plane, though some points sit above $C_{\text{root}} = 0.25$. Seeded starts widen the front toward higher $M_\alpha$ and $N_{wall}$ up to about $5\text{ MW}/\text{m}^{-2}$, and keep the reported Pareto set below $C_{\text{root}} = 0.25$. The random to seeded gain for YF-3 is moderate. It is smaller than the corresponding gain seen for YF-2 in Fig.~\ref{fig:Pareto_yf2}.

\par In short, YF-2 returns a $Q_{phys} \sim 1$ burning point with $C_{\text{root}} = 0.20$ and $P_{aux} = 13.6 \text{ MW}$. YF-3 returns an ignited reactor scale point with $C_{\text{root}} = 0.13$ and $M_\alpha = 3.3$. Both equal weight solutions lie in the electron root favorable band of Sec.~\ref{sec:2.6_ele_root_proxy}. Seeding helps both stages. The improvement is clearer for YF-2 than for YF-3.

\section{Conclusions}
\label{sec:5_conclusion}
\par We developed a 0D physics model and a multiphysics constrained optimization framework for SQuIDs. The framework couples confinement scaling, density limits, radiation, fusion gain and EC/EBW heating accessibility, together with an electron root proxy, and performs single objective and multi objective searches for YF-1, YF-2 and YF-3 according to each device stage.

\par 0D scans show that density peaking $S_n$ reshapes EC/EBW access at fixed volume averaged density. Larger $S_n$ raises the central density $n_{e0}$, so more operating points exceed the O1/X1 cutoffs and electromagnetic access shrinks. The EBW channel targets this overdense interval; the same rise in $n_{e0}$, with a shorter density scale length near the conversion layer, therefore expands the EBW accessible window within the present cold plasma bracket.

\par For YF-1, minimising $P_{aux} = 0.32\text{MW}$ yields $R_0 = 2.26\text{ m}$, $B_0 = 1.64\text{ T}$, $A = 11.00$. For YF-2, we obtain $R_0 = 6.00\text{ m}$, $B_0 = 4.70\text{ T}$, $A = 3.00$, $Q_{phys} = 1.05$, $P_{aux} = 13.6\text{ MW}$ and $C_{\text{root}} = 0.20$. For YF-3, the equal weight point is ignited at reactor scale: $R_0 = 17.00\text{ m}$, $B_0 = 9.30\text{ T}$, $A = 5.00$, $P_{fus} = 5.16\text{ GW}$, $M_\alpha = 3.3$ and $C_{\text{root}} = 0.13$. Both YF-2 and YF-3 optima lie in the electron root favorable band.

\par These 0D results are indicative. Future work will couple engineering and economic modules and benchmark against systems codes and higher fidelity calculations.

\section*{Appendix A. EBW heating feasibility via O--SX--B mode conversion}
\label{sec:appendix_a}
\par RF power may be coupled to EBWs through O--SX--B mode conversion in high density stellarator plasmas. Launch enters through $N_z = k_z / k_0$ with $N_y = 0$, consequently, coupling efficiency depends on a single launch angle in the $B_0$--$k$ plane under the cold plasma WKB model below.

\par The O-mode cutoff density is determined by $B_0$, expressed as:
\begin{subequations}
\begin{align}
\omega_{gyro} &= \min(h \omega_{ce}(B_0),\, \omega_{\max}), \label{eq:A1a} \\
n_{O,cut} &= \frac{m_e \varepsilon_0}{e^2} \omega_{gyro}^2, \label{eq:A1b}
\end{align}
\end{subequations}
where h = 1 (O1) or 2 (X2) according to the chosen cutoff mode. Evaluating the main text plasma profiles at the O-mode cutoff radius $r_{cut}$ yields the characteristic scale length $L_n \equiv L_n(r_{cut})$, defined via:
\begin{subequations}
\begin{align}
r_{cut} &= a \sqrt{1 - \left( \frac{n_{O,cut}}{n_{e0}} \right)^{\frac{1}{S_n}}}, \label{eq:A2a} \\
L_n(r) &= \frac{(1 - x^2) a^2}{2 S_n r}, \quad x = \frac{r}{a}. \label{eq:A2b}
\end{align}
\end{subequations}

\par We then define the normalized frequency $Y_{cut} = \frac{\omega_{ce}(r_{cut})}{\omega}$, the toroidal field profile $B(r) = \frac{B_0 R_0}{R_0 + r}$, and the magnetic scale length $L_B = R_0$. Cold plasma upper hybrid resonance (UHR) position $r_{UHR}$, naturally satisfies the local resonance condition $\omega_{pe}^2(r) + \omega_{ce}^2(r) = \omega^2$.

\par Under optimal launch conditions, the transmission coefficient for the O-SX conversion stage is governed by:
\begin{equation}\tag{A3}\label{eq:A3}
N_{z,opt} = \sqrt{\frac{Y}{Y + 1}},
\end{equation}
hereafter $Y \equiv Y_{cut} = \omega _{ce}(r_{cut})/\omega $,
\begin{equation}\tag{A4}\label{eq:A4}
T_{O-SX} = \exp\left\{ -\pi k_0 L_n(r_{cut}) \times \sqrt{\frac{Y}{2}} \left[ 2(1 + Y)\left(N_z - N_{z,opt}\right)^2 + N_y^2 \right] \right\},
\end{equation}
where $k_0 = \omega/c$ is the vacuum wavenumber.

\par Subsequent SX-B mode conversion occurring at the UHR layer yields an efficiency evaluated as:
\begin{equation}\tag{A5}\label{eq:A5}
C_{\text{SX}-B} = 1 - \exp(-\pi\eta),
\end{equation}
where the optical depth parameter $\eta$ is explicitly derived as:
\begin{equation}\tag{A6}\label{eq:A6}
\eta = \frac{k_0 L_{n,UHR} \left(1 + N_z^2\right)}{\left(1 + \alpha^2\right) \left(2 - \alpha^2 N_z^2 - \dfrac{\alpha^2 N_z^2}{1 + N_z^2}\right)}.
\end{equation}

\par Overall, the total coupling efficiency $\eta_{O-SX-B}$ is established by the product of the individual stage efficiencies:
\begin{equation}\tag{A7}\label{eq:A7}
\eta_{O-SX-B}\left(N_z\right) = T_{O-SX}\left(N_z\right) C_{\text{SX}-B}\left(N_z\right).
\end{equation}
Maximizing $\eta_{O-SX-B}\left(N_z\right)$ over the accessible injection window $N_z \in [0.3, 1.2]$ dynamically determines $\eta_{O-SX-B,\max}$ and the corresponding optimal refractive index $N_{z,opt}$ for each distinctive operating regime.

\par Conclusively, EBW heating is feasible if and only if the plasma parameters satisfy the following coupled optimization constraints:
\begin{equation}\tag{A8}\label{eq:A8}
\begin{cases}
B_0 \le B_{\max} \\[4pt]
n_{O,cut} < n_{e0} < n_{hi} \\[4pt]
\dfrac{r_{cut}}{a} < f_{cut} \\[10pt]
\dfrac{r_{UHR}}{a} < f_{UHR} \\[10pt]
L_n\left(r_{cut}\right) > 0 \\[4pt]
\eta_{O-SX-B,\max} \ge \eta_{\min}
\end{cases},
\end{equation}
where $f_{cut} = 0.92$ and $f_{UHR} = 0.98$ are the maximum allowed normalized radii $r_{cut}/a$ and $r_{UHR}/a$, and $\eta_{\min} = 0.10$ is the minimum acceptable coupling efficiency. Here, $n_{hi}$ is obtained by increasing $n_{e0}$ until one of the radial constraints in Eq.~(\ref{eq:A8}) violated. Below the electromagnetic cutoff the design relies on O/X-mode EC access; the EBW branch is evaluated in the overdense window $n_{O,cut} < n_{e0} < n_{hi}$ via cold-plasma O--SX--B conversion.

\par Note that Eq.~(\ref{eq:A3})--(\ref{eq:A7}) follow Guo et al.~\cite{27guo2017one} ($k_0 L_n \gg 1$). We particularly emphasize that this cold plasma slab model brackets accessible heating regions in the 0D scan; it does not resolve antenna launch, ray tracing, poloidal ripple of $\boldsymbol{B}$ and $n_e$, or full-wave mode conversion beyond WKB limit.

\bibliographystyle{iopart-num}
\bibliography{references}     

\end{document}